\documentclass[10pt,twocolumn]{article}

\usepackage{graphicx}
\usepackage{amsfonts}
\usepackage{amsmath}
\usepackage{amssymb}
\usepackage{mathrsfs} 
\usepackage{lipsum}
\usepackage{color}
\usepackage[squaren,Gray,cdot]{SIunits}
\usepackage{authblk}
\usepackage[top=2cm, bottom=2cm, left=2cm, right=2cm]{geometry}
\usepackage{mathtools, cuted}
\usepackage{array,amsmath,booktabs}
\usepackage[font=small, labelfont={bf,normalsize}]{caption}

\usepackage{sectsty}
\sectionfont{\sffamily}
\subsectionfont{\sffamily}
\subsubsectionfont{\sffamily}
\paragraphfont{\sffamily}

\newcommand{\Hsp}{H_{\mathrm{sp}}}
\newcommand{\perfluoro}{\mathrm{C}_6 \mathrm{F}_{14}}
\newcommand{\fres}[1]{f_{\mathrm{res},#1}}

\newcommand{\cAapp}{c_{\mathrm{out}}}
\newcommand{\cA}{c_{\mathrm{a}}}
\newcommand{\phis}{\varphi_s}
\newcommand{\Rpore}{R_{\mathrm{eq}}}
\newcommand{\alphapore}{\alpha_{\mathrm{v}}}
\newcommand{\alphath}{\alpha_{\mathrm{th}}}
\newcommand{\lth}{\ell_{\mathrm{th}}}
\newcommand{\lv}{\ell_{\mathrm{v}}}
\newcommand{\hM}{h_{\mathrm{M}}}

\begin{document}
%
%
\title{\sf Ultrasound transmission through monodisperse 2D microfoams}
\author[1]{Lor\`ene Champougny}
\author[2]{Juliette Pierre}
\author[1]{Antoine Devulder}
\author[3]{Valentin Leroy}
\author[1]{Marie-Caroline Jullien}
\affil[1]{\normalsize{Gulliver, CNRS, ESPCI Paris, PSL University, 10 rue Vauquelin, 75005 Paris, France.}}
\affil[2]{\normalsize{Institut Jean Le Rond d'Alembert, CNRS, Sorbonne Universit\'es, UPMC Univ. Paris 6, Paris, France.}}
\affil[3]{\normalsize{Laboratoire Mati\`ere et Syst\`emes Complexes, CNRS, Universit\'e Paris-Diderot, Sorbonne Paris Cit\'e, Paris, France.}}
\date{}
%
%
\twocolumn[
\begin{@twocolumnfalse}

\maketitle
           
\begin{abstract}
While the acoustic properties of solid foams have been abundantly characterized, sound propagation in liquid foams remains poorly understood.
Recent studies have investigated the transmission of ultrasound through three-dimensional polydisperse liquid foams (Pierre \textit{et al.}, 2013, 2014, 2017).
However, further progress requires to characterize the acoustic response of better controlled foam structures.
In this work, we study experimentally the transmission of ultrasounds through a single layer of monodisperse bubbles generated by microfluidics techniques.
In such a material, we show that the sound velocity is only sensitive to the gas phase. 
Nevertheless, the structure of the liquid network has to be taken into account through a transfer parameter analogous to the one in a layer of porous material.
Finally, we observe that the attenuation cannot be explained by thermal dissipation alone, but is compatible with viscous dissipation in the gas pores of the monolayer.
\end{abstract}

\vspace{0.5cm}
\end{@twocolumnfalse}]
%
\section{Introduction} \label{intro}
%
%
Foams consist in dense assemblies of gas inclusions in a condensed phase matrix -- either solid or liquid.
Such materials are known for their sound-attenuating properties: solid foams are widely used for soundproofing purposes in buildings \cite{Attenborough1982}, while liquid foams can efficiently mitigate explosions \cite{Raspet1983}.
However, the acoustical properties of the latter remain comparatively less investigated and understood.
%

Recent works have shed light on the sound propagation mechanisms within liquid foams, following two distinct pathways.
On the one hand, studies on macroscopic polydisperse foam samples have demonstrated the highly dispersive behavior of liquid foams at ultrasonic frequencies, due to the mechanical coupling between the liquid network and the thin films separating the bubbles \cite{Pierre2013,Pierre2014}.
Some studies have identified thermal losses as the main dissipation source in liquid foams \cite{Mujica2002,Monloubou2015}, while another work evidenced an additional dissipation mechanism at low frequencies \cite{Pierre2017}.
However, the bulk structure of those 3D samples is not known precisely, and assumptions have to be made in order to relate the foam structure to its acoustical properties.
On the other hand, the acoustic responses of the foams' constitutive elements, namely single thin liquid films \cite{Kosgodagan2014} and liquid channels \cite{Derec2015}, have also been investigated separately at low frequencies.
However, collective effects cannot be captured in those individual units.
Thus there is a need for model systems of intermediate complexity, allowing collective effects to emerge while keeping a structure simple enough to be accurately characterized.

In the last decades, microfluidics has proven to be a powerful tool to generate monodisperse bubble assemblies \cite{Anna2003,Garstecki2004} and manipulate them in a controlled way \cite{Marmottant2009,Raven_PhD}.
Monodisperse (quasi-) 2D foams consist in a single layer of identical bubbles confined between two surfaces.
Thanks to their simpler structure \cite{Cox2008,Gay2011} compared to 3D foams, they have become widespread model systems \cite{Drenckhan2010}, used in particular at microscale to explore questions related to foam rheology or controlled drainage \cite{Miralles2014,Huerre2014}.
However, to the best of our knowledge, the acoustic properties of 2D microfoams (or bubble monolayers) have never been investigated in the literature. 

In this work, we propose to contribute to the understanding of the relationship between the acoustic response of a liquid foam and its structure by studying the transmission of ultrasound through a model system of intermediate complexity: a microfluidics-generated monodisperse 2D microfoam.
In section \ref{sec:exp}, we present our experimental setup, which includes an original microfluidic cell designed specially for ultrasonic measurements through low acoustic impedance samples. 
Section \ref{sec:results} then reports raw experimental data for the acoustic transmission (in the range $100 - 1000~\kilo\hertz$) through our cell, either filled only with air or with 2D microfoams of various liquid fractions and bubble sizes.
In section \ref{sec:analysis}, we present a model inspired from the literature on porous materials, allowing us to analyze the transmission curves.
The sound velocity and attenuation within the monolayers are extracted from the analysis and discussed.
Our conclusions are summerized in section \ref{sec:conclusion}.
%
%
\section{Materials and methods}\label{sec:exp}
%
In this section, we describe the fabrication of the microfluidic cell in which our samples will be studied (paragraph \ref{ssec:cell}), the microfoams' generation and characterization (paragraphs \ref{ssec:generation} and \ref{ssec:mono_charac}), and finally the acoustic setup and protocole (paragraph \ref{ssec:protocole_acou}).
%
\subsection{Design and fabrication of the cell} \label{ssec:cell}
\begin{figure}[t!]
\centering
\includegraphics[width=7cm]{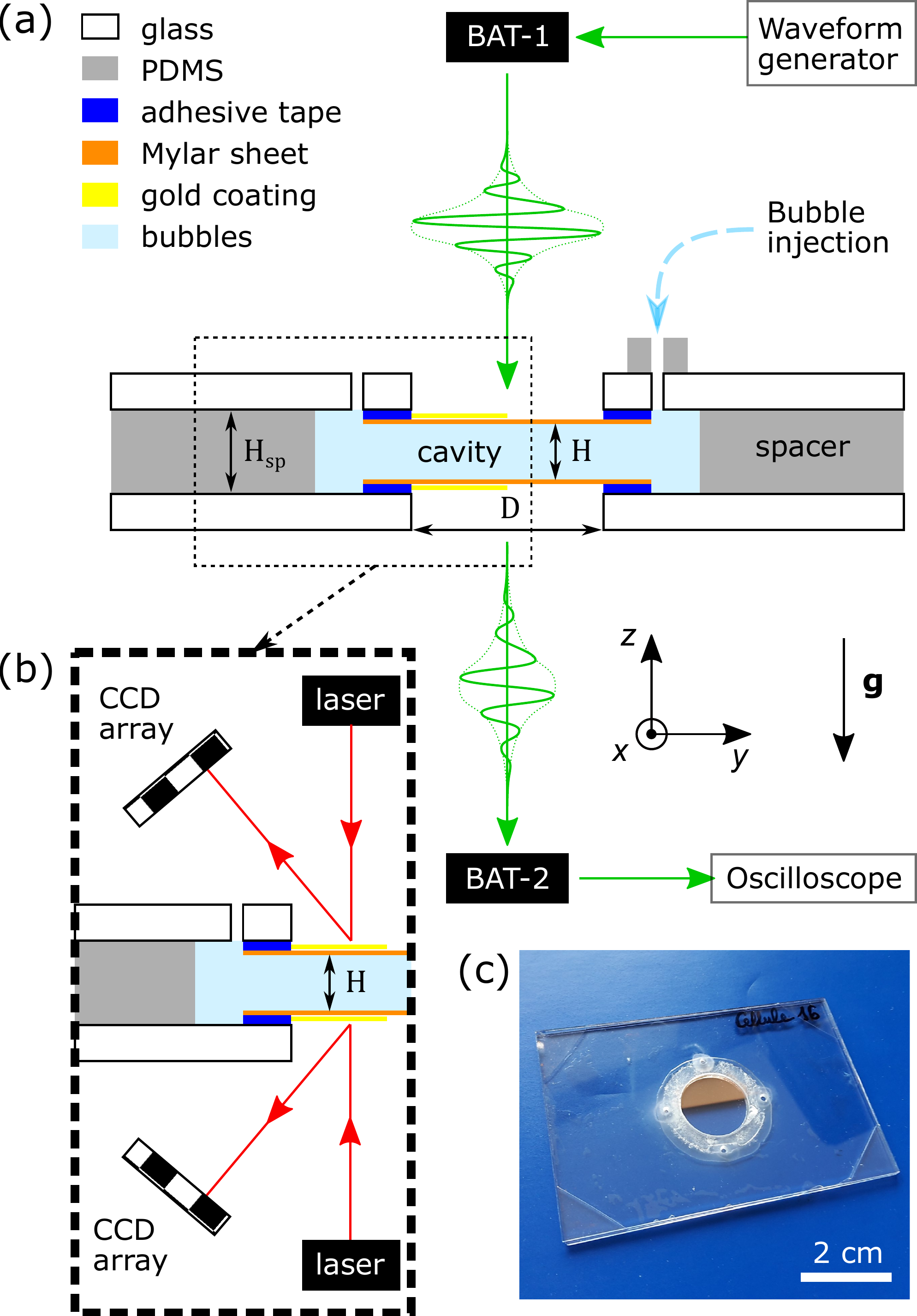}
\caption{(a) Sketch of the simplified ultrasonic setup and microfluidic cell designed for ultrasonic transmission measurements through low acoustic impedance samples, including foams. The relative scales of the elements have been changed in order to ease reading. --- (b) Sketch of the laser telemetry setup used to measure the cavity thickness $H$. --- (c) Top photograph of the microfluidic cell.}
\label{fig:cell}
\end{figure}
When designing the cell, the main challenge was to comply with the requirements set by the length scales of both airborne ultrasounds (typically millimetric) and microfluidics (typically micrometric).
A side view of our microfluidic cell is depicted in Fig. \ref{fig:cell}a.
It consists of two glass slides in which a circular hole of diameter $D = 16~\milli\meter$ is made using a laser cutting machine.
Four small additional holes (less than $1~\milli\meter$ in diameter) are pierced in one of the slides, at the periphery of the main hole, in order to serve as inlets and outlets.
The main $16~\milli\meter$ holes are covered by thin PET films (polyethylene terephtalate, also referred to as Mylar), attached by the way of laser-cut double-sided adhesive tape annuli.
The thickness of the Mylar membranes ($\hM \approx 1~\micro\meter$) is chosen as small as possible in order to let airborne ultrasound through.

The spacing between the two glass slides is set by a crosslinked PDMS spacer (poly(dimethyl siloxane), RTV purchased from Momentive).
This spacer is fabricated by spin-coating and then baking successive PDMS layers in order to reach approximately the desired thickness.
The actual thickness $\Hsp$ of the resulting spacer is determined by optical profilometry.
The cavity that will contain the bubbles is created by cutting a circular hole of diameter larger than $D$ in the spacer.
The cell is then assembled and the cavity is sealed by oxygen plasma-bonding the glass slides on both sides of the PDMS spacer.
This plasma treatment also has the advantage of making the inside of the cavity hydrophilic, which is necessary to work with aqueous solutions.
In the experiments shown in section \ref{sec:results}, we used two different cells, named C1 and C2.
The associated spacer thicknesses $\Hsp$, empty cavity thicknesses $H$ and Mylar membrane thicknesses $h_M$ are given in table \ref{tab:cavities}.
\begin{table}
\centering
\begin{tabular}{cccc}
 Cavity & $\Hsp~(\micro\meter)$ & $H~(\micro\meter)$ & $\hM~(\micro\meter)$ \\
\hline
C1 & $376 \pm 3$ & $212 \pm 3$ & $0.97 \pm 0.01$ \\
C2 & $483 \pm 4$ & $312 \pm 7$ & $1.03 \pm 0.02$ \\
\end{tabular}
\caption{Thicknesses of the spacer $\Hsp$, cavity $H$ (when empty) and Mylar membrane $\hM$ (after metallization) for the cells C1 and C2 used in the experiments.}
\label{tab:cavities}
\end{table}
%
%
\subsection{Sample preparation} \label{ssec:generation}
%
\emph{Foaming solution.}
The liquid phase of the foam is an aqueous solution of sodium dodecyl sulfate (SDS, purchased from Sigma Aldrich) at a concentration of $2.5~\gram\per\liter$, corresponding to about $1.1$ times the critical micellar concentration. 
Some glycerol (purchased from Sigma Aldrich) is added at a concentration of $5~\%$ in weight in order to slow down film drainage. \bigskip \\ 
%
\emph{Gas composition.}
The gaseous phase consists in a mixture of air and perfluorohexane vapor ($\perfluoro$, purchased in liquid state from Alfa Aesar) with a volume fraction of $0.22 \pm 0.03$.
The mixture properties were measured using the acoustic transmission through a cavity filled with gas phase only (data not shown). 
The density and sound velocity of the mixture were found to be $\rho_g = 3.9 \pm 0.6~\kilo\gram\per\meter^3$ and $c_g = 184 \pm 7~\meter\per\second$, respectively.

Due to its poor solubility in water, perfluorohexane slows down gas transfer between adjacent bubbles, a phenomenon known as Oswald ripening \cite{Cantat2013}, which would otherwise occur within minutes in the microfoam \cite{Marchalot2008}.
With this gas mixture, we observe that the bubble size remains constant during the first hour after foam generation.
Hereafter, our measurements are always performed within this time lapse. \bigskip \\
%
\emph{Bubble generation.}
Monodisperse bubbles are generated using a classical microfluidic flow-focusing junction \cite{Anna2003,Garstecki2004}, in which the gas phase is being pinched on both sides by the liquid phase when entering a constriction.
The phases are set into motion by applying an overpressure at the inlets with a pressure controller (Fluigent, MFCS-EZ).
The bubble size can be tuned by changing the geometry of the constriction or the pressure applied to the phases \cite{Garstecki2004,Raven_PhD}.
As they are being produced, the bubbles are injected into the measurement cavity and pack together to form a monolayer.
%
\subsection{Sample characterization} \label{ssec:mono_charac}
\begin{figure}[t!]
\centering
\includegraphics[width=7.5cm]{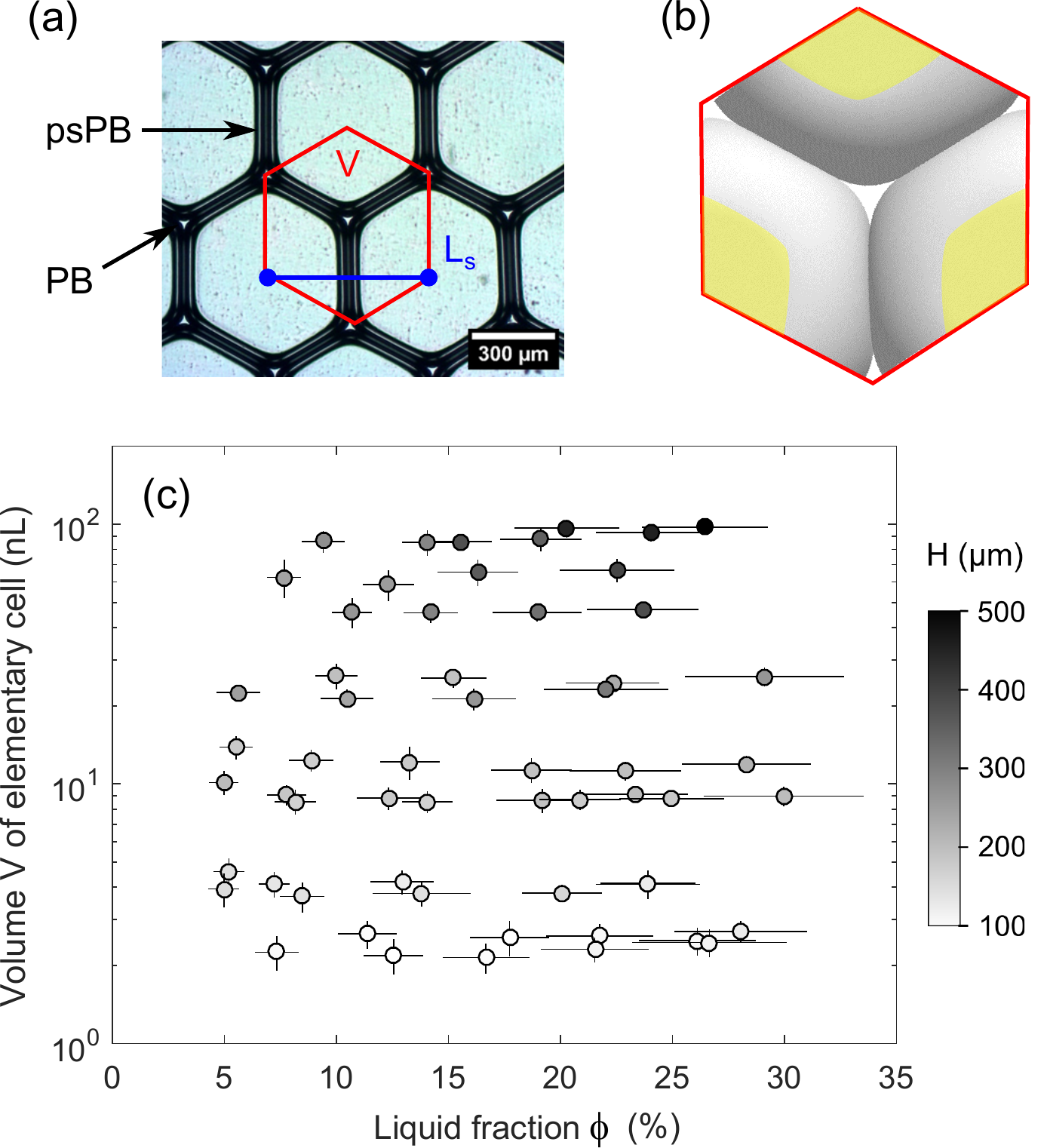}
\caption{(a) Microscope picture of a monolayer showing the definition of the hexagonal elementary cell of volume $V$, the bubble center-to-center distance $L_s$, the Plateau borders (PB) and pseudo Plateau borders (psPB). --- (b) Top view of a simulated elementary cell for a monolayer with $\phi = 15~\%$ and $H / L_s = 1$. The surfaces wetting the top plate are highlighted in yellow. --- (c) Liquid volume fraction $\phi$ and hexagonal cell volume $V$ for all the bubble monolayers investigated in our experiments. The grey level codes for the measured cavity thickness $H$.}
\label{fig:summary}
\end{figure}
%
\emph{Cavity thickness.}
An important feature of our setup is that, albeit not elastic, the $1~\micro\meter$-thick Mylar sheets that close the cavity are flexible.
Consequently, the cavity can deform depending on the foam sample it encloses.
However, as developed in Appendix \ref{apdx:profile}, the topography of the Mylar sheets shows no large-scale variations in the presence of a monolayer.
Thus, the thickness $H$ remains homogeneous over the whole cavity, but still has to be determined independently.

For each experiment, the cavity thickness is measured by the means of a laser telemetry setup (L-LAS-LT-MS-37, Sensor Instruments), exploiting the reflection of laser beams ($1~\milli\watt$ at most, $\lambda = 670~\nano\meter$) on the Mylar sheets, as sketched in Figure~\ref{fig:cell}b.
The overall setup yields a typical uncertainty of $\pm 10~\micro\meter$ on the cavity thickness. 
In order to ensure a good reflectivity, half the surface of the Mylar membranes is coated with a gold layer of a few tens of $\nano\meter$ in thickness, while the other half remains transparent for sample imaging purposes (see Fig. \ref{fig:cell}c).
Due to both manufacturing variability and uncertainties in the coated gold thickness, the exact thickness of the metallized membrane $\hM$ will be determined in paragraph \ref{ssec:empty_cavity} for cavities C1 and C2.
\bigskip \\
%
\emph{Bubble size.}
The bubble monolayer is observed with an optical microscope (DMI 6000B, Leica) at magnifications $\times 2.5$ or $\times 5$, depending on the bubble size.
The monodisperse bubbles spontaneously organize into a regular hexagonal lattice, as shown in Fig.~\ref{fig:summary}a. 
The bubble center-to-center distance $L_s$ can be computed from the analysis of pictures taken with a CCD camera (PixeLink).
However, since $L_s$ varies depending on the cavity thickness, we rather choose the volume $V = \sqrt{3}/2 \times L_s^2 H$ of the elementary hexagonal cell as a measurement of the bubble size. \bigskip \\
%
\emph{Liquid fraction.}
The liquid fraction $\phi$ of the monolayer is defined as the ratio between the volume of liquid within the foam and the total volume of the foam.
Right after production, the liquid fraction of the monolayer usually reaches a value around $30\%$ in our experimental conditions.
For a given monolayer, $\phi$ can be varied by carefully absorbing some liquid with a tissue put in contact with one of the cavity inlets.
The actual value of the liquid fraction is measured by weighting the sample with a high-precision scale (TE 64, Sartorius). 
Knowing the mass of the empty cell, the volume of the cavity and the density of the liquid phase, we can deduce $\phi$ by neglecting the weight of the gas phase.
This method yields a typical relative error of $\pm 0.1$ on the liquid fraction. \bigskip \\
%
\emph{Surface pore fraction.}
We define the surface pore fraction $\phis$ as the fraction of Mylar walls' surface that is not covered by the liquid network (PB and psPB).
Qualitatively, $\phis$ roughly corresponds to the bright area fraction in Fig.~\ref{fig:summary}a, but one has to keep in mind that the black pattern corresponds to total reflection zones and not to the PB and psPB network directly \cite{vandeNet2007b,Gaillard2015}.

In order to retrieve the surface pore fraction $\phis$, we determine the monolayer real structure using the surface minimization code Surface Evolver, developed by K. Brakke \cite{Brakke1992}.
For each monolayer, the structure is computed knowing the liquid fraction $\phi$ and the ratio $H/L_s$, both of which are measured in the experiments.
In the simulations, once the monolayer has converged towards its equilibrium shape, the wetted area is obtained by extracting the surface elements in contact with the cavity wall (colored area in Fig.~\ref{fig:summary}b) and $\phis$ is deduced. \bigskip \\
%
%
\emph{Overview of samples.}
As summarized in Fig.~\ref{fig:summary}c, a total of 58 samples was tested in our experiments.
Liquid fractions varied in the range $5~\% \lesssim \phi \lesssim 30~\%$ and bubble center-to-center distances in the range $150~\micro\meter \lesssim L_s \lesssim 600~\micro\meter$.
The cavity thicknesses, which lied in the range $100~\micro\meter \lesssim H \lesssim 500~\micro\meter$, are coded by the greyscale in Fig.~\ref{fig:summary}c, going from lighter to darker symbols.
The bubble volumes deduced from the measurements of $L_s$ and $H$, were found in the range $2~\nano\liter \lesssim V \lesssim 100~\nano\liter$.
It can be noted that the thickness $H$ of the (flexible) cavity increases with both liquid fraction and bubble size.
%
\subsection{Acoustic measurements} \label{ssec:protocole_acou}
%
\emph{Ultrasonic setup.}
The ultrasonic setup used in our experiments is very close to the one introduced by Pierre \textit{et al.} \cite{Pierre2013}. It is specifically designed for the measurement of ultrasound transmission through low-acoustic-impedance samples, such as porous materials with a large gas content.
Gaussian-shaped pulses are generated using a waveform generator (Handyscope HS5, Tiepie), further amplified by a power amplifier (WMA-300, Falco Systems) and converted into ultrasonic airborne pulses by a broadband air transducer (BAT-1 source, MicroAcoustic). After propagation through the sample of interest, enclosed in the specially designed cell (see Fig.\ref{fig:cell}a), the pulses are received by another transducer (BAT-2 receiver, MicroAcoustic), amplified again by a pre-amplifier (Olympus NDT, $40~\deci\bel$) and recorded by a digital oscilloscope (Handyscope HS5, Tiepie). \bigskip \\
%
\emph{Data acquisition and processing.}
In a typical experiment, three Gaussian pulses, centered at $150$, $400$ and $800~\kilo\hertz$ respectively, are sent through the sample, allowing to measure the frequency-dependence of the transmission across a decade ($100$ to $1000~\kilo\hertz$).
The acoustic time signals are recorded with a sampling frequency of $20~\mega\hertz$ and averaged over ten successive acquisitions in order to improve the signal to noise ratio.
For each central frequency (\textit{i.e} $150$, $400$ or $800~\kilo\hertz$), two different pulses are recorded.
First, a reference pulse REF is acquired when the sample is replaced by a simple glass slide pierced with a hole of the same diameter $D$ as the cell.
In this situation, the acoustic wave propagates only through air but diffraction effects, due to the finite hole size compared to the wavelength, are accounted for.
Then, the cell is put on the path of the acoustic wave and the SAM pulse, that has propagated through the sample, is recorded.
Our observable is the complex transmission $T$ through the sample, defined as the ratio between the respective Fourier transforms $T_{\mathrm{SAM}}$ and $T_{\mathrm{REF}}$ of the SAM and REF signals
\begin{equation}
T(f) = \frac{T_{\mathrm{SAM}}(f)}{T_{\mathrm{REF}}(f)},
\end{equation}
which is a function of the frequency $f$.
\begin{figure}[t!]
\centering
\includegraphics[width=7.5cm]{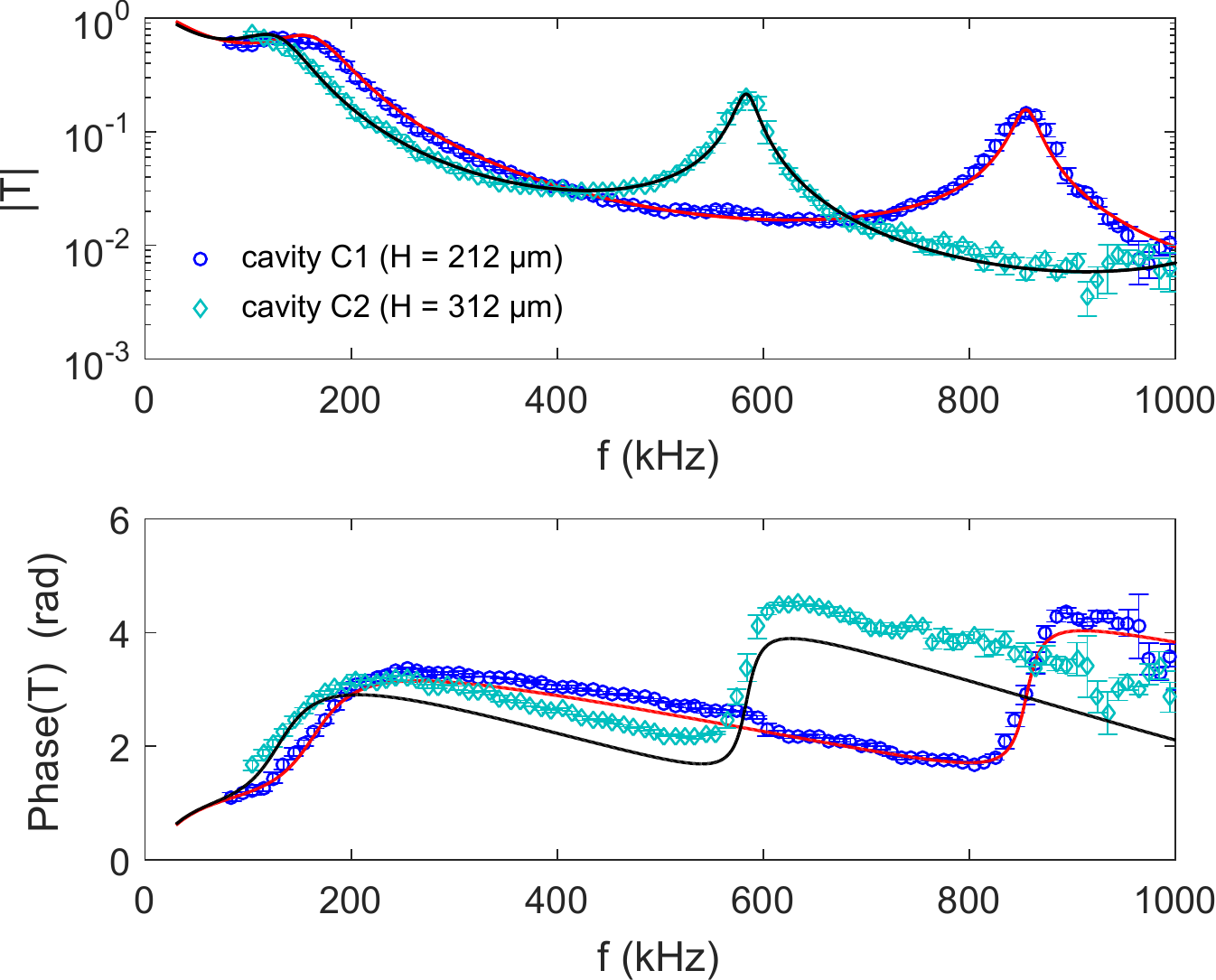}
\caption{Modulus and phase of the acoustic transmission $T$ through cavities C1 and C2 filled with air, as a function of the frequency $f$ (symbols). The solid lines are fits using the Mylar membrane thickness $\hM$, sound velocity $\cA$ and attenuation $\alpha$ in air as adjustable parameters. The fit values are $\hM=0.97~\micro\meter$, $\cA=349~\meter\per\second$ and $\alpha=180~\meter^{-1}$  for C1 and $\hM=1.03~\micro\meter$, $\cA=346~\meter\per\second$ and $\alpha=154~\meter^{-1}$ for C2.}
\label{fig:transmission_air}
\end{figure}
%
\section{Experimental results} \label{sec:results}
%
\subsection{Transmission through air-filled cavity} \label{ssec:empty_cavity}
Before investigating the transmission through a bubble-filled cavity, we first characterize the acoustic response of our microfluidic cells when they only contain air.
The symbols in Fig.~\ref{fig:transmission_air} show the modulus and phase of the complex transmission measured through cavities C1 and C2 filled with air at atmospheric pressure and room temperature. 
As observed by Pierre \textit{et al.} \cite{Pierre2013} on similar systems, the cavity behaves as an acoustic Fabry-P\'erot interferometer \cite{Hernandez1988}, hence the resonant behavior observed on the transmission in Fig.~\ref{fig:transmission_air}.
The $n^{\mathrm{th}}$ resonance frequency $\fres{n}$ is given by
\begin{equation}
\fres{n} = n \times \frac{c}{2H},
\label{eq:resonance}
\end{equation}
where $c$ is the velocity of sound in the medium enclosed in the cavity (air in this example) and $n$ is a positive integer.

The transmission through the five-layer system consisting of air / Mylar / medium in cavity / Mylar / air can be computed exactly knowing the thickness, acoustic impedance and wave vector of each medium \cite{Brekhovskikh2012,Pierre2013}. 
In Figure~\ref{fig:transmission_air}, the solid lines are fits of the data using the wave vector in air $k_a = 2\pi f/\cA - i\alpha$ and the Mylar membrane thickness $\hM$ as adjustable parameters. 
For both cavities, the sound velocity $\cA$ consistently lies between $346$ and $349~\meter\per\second$. 
The adjusted membrane thicknesses $\hM$ are displayed in table \ref{tab:cavities}.
These values allow to account for the mass added to the membrane by metalization and will be used in the data analysis (section \ref{sec:analysis}).
%
%
\subsection{Transmission through monolayers} \label{ssec:raw_data}
\begin{figure}[t!]
\centering
\includegraphics[width=8cm]{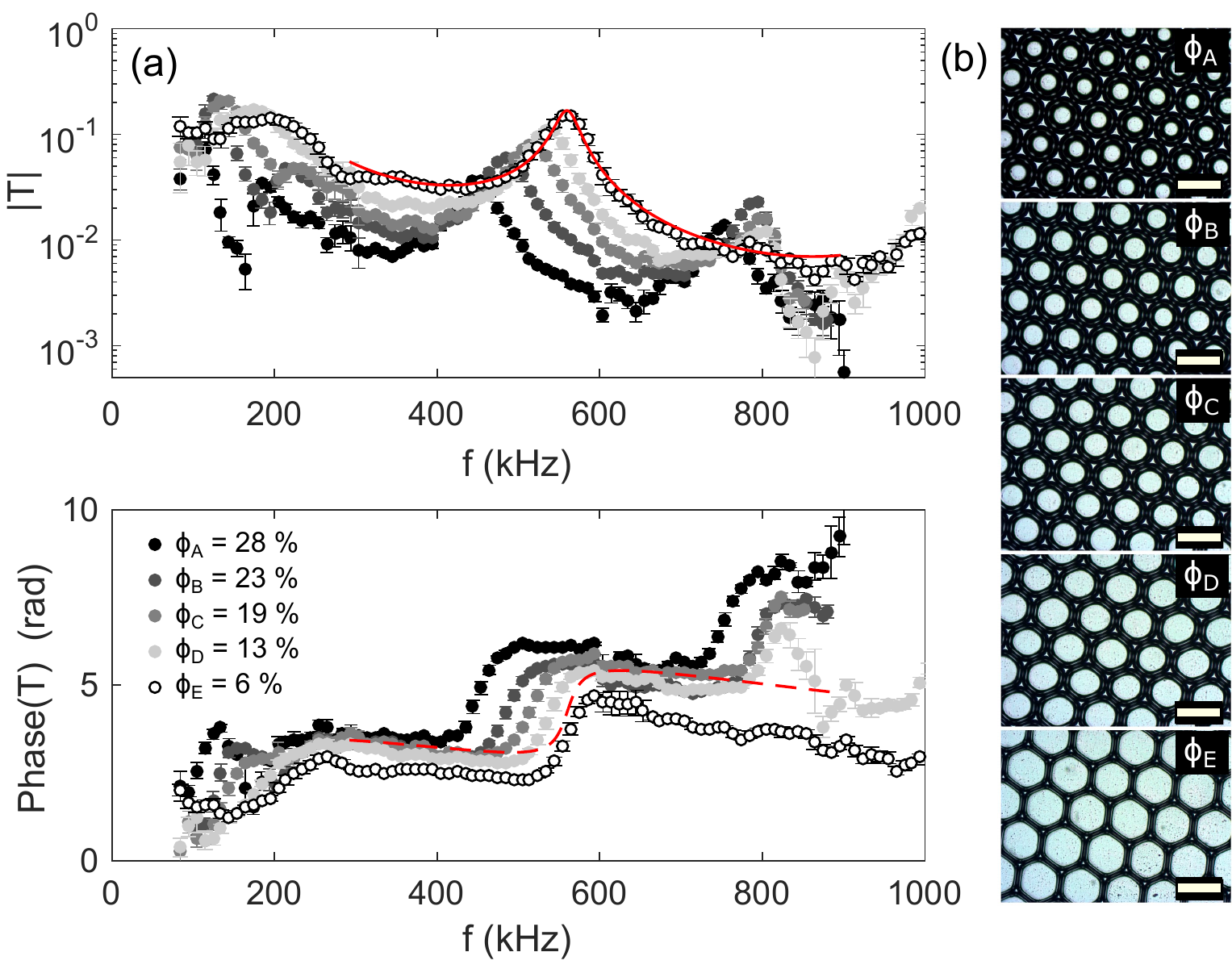}
\caption{(a) Modulus and phase of the acoustic transmission $T$ as a function of the frequency $f$ for a cavity filled with bubble monolayers of fixed hexagonal cell volume $V = 12~\nano\liter$ and various liquid fractions $\phi$ (symbols). The solid red line is a fit of the data for $\phi_E$ using the model described in section \ref{ssec:sieve}. --- (b) Pictures of the corresponding monolayers. The scalebar represents $300~\micro\meter$.} 
\label{fig:transmission_mono}
\end{figure}
\begin{figure}[t!]
\centering
\includegraphics[width=8cm]{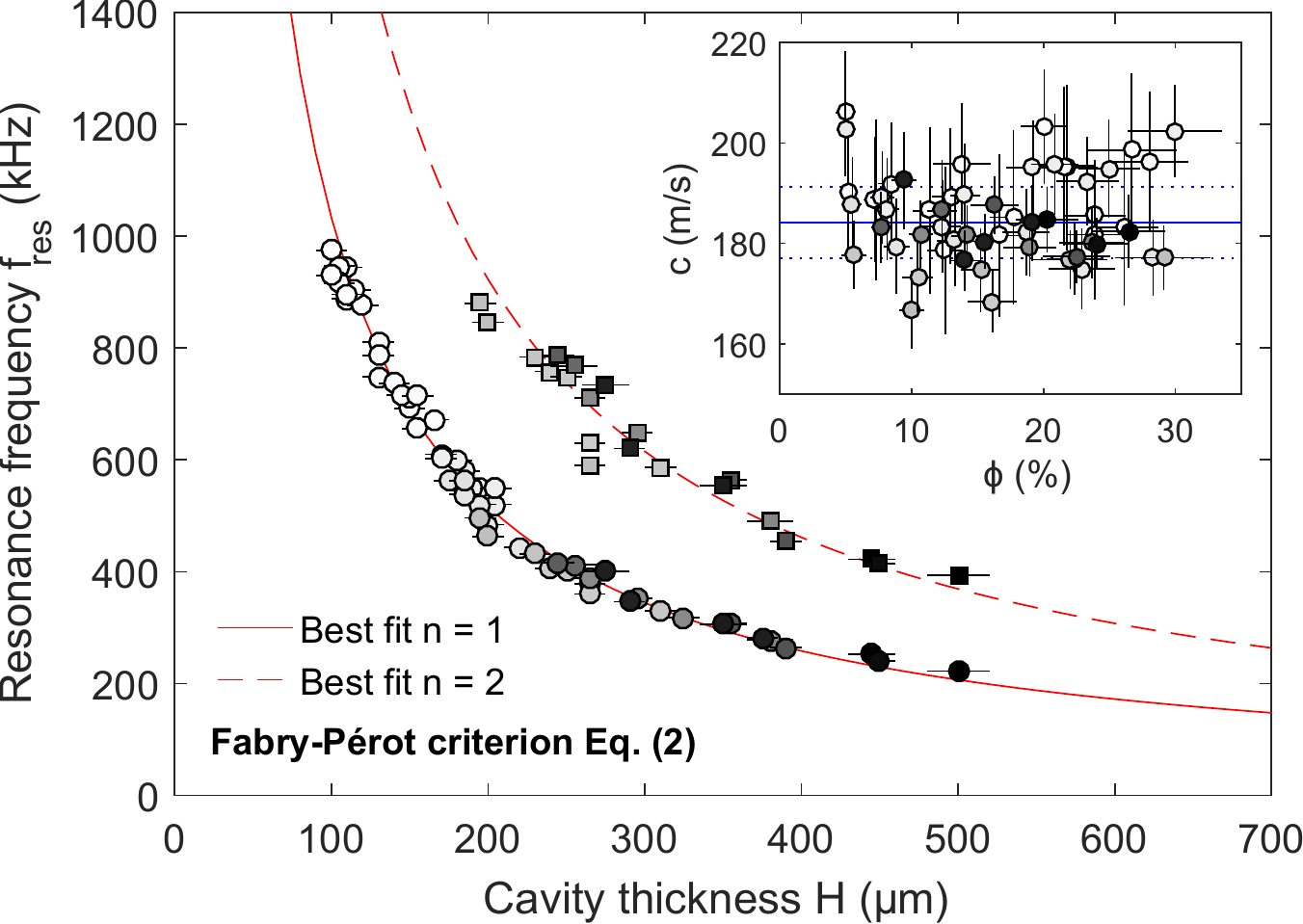}
\caption{Main plot: frequencies of the first ($\circ$) and second ($\square$) resonances in bubble monolayers as a function of the measured cavity thickness $H$. The solid and dashed lines show the Fabry-P\'erot resonance criterion Eq. \eqref{eq:resonance} for $n=1$ and $2$, with the best-fitting sound velocities $207~\meter\per\second$ and $185~\meter\per\second$, respectively. --- Inset: sound velocity $c$ in the cavity, extracted from the analysis described in section \ref{sec:analysis}, as a function of the liquid fraction $\phi$. The solid line shows the sound velocity in the gas phase $c_g = 184~\meter\per\second$, measured independently. Dotted lines show the confidence interval $\pm 7~\meter\per\second$ on $c_g$. In both plots, the greyscale codes for the bubble volume $V$ (the darker the symbol, the larger the bubble volume).}
\label{fig:Fres}
\end{figure}
As an illustration of the transmission data we obtain when bubble monolayers are enclosed in the cavity, Fig.~\ref{fig:transmission_mono}a shows the complex acoustic transmission $T$ through monolayers of constant hexagonal cell volume $V = 12~\nano\liter$ and various liquid fractions. 
Photographs of the corresponding structures are displayed in Fig.~\ref{fig:transmission_mono}b.
Contrary to the case of 3D liquid foams \cite{Pierre2013}, the attenuation in 2D foams is small enough so that a Fabry-P\'erot resonance may still be observed in the transmission curves.
As can be seen in Fig.~\ref{fig:transmission_mono}a, the resonance frequency seems to vary with the liquid fraction $\phi$.
However, we recall that the cavity thickness $H$ changes depending on the monolayer enclosed, and is measured for each sample.
For instance, in the data presented in Fig.~\ref{fig:transmission_mono}a, $H$ varies from $220~\micro\meter$ for $\phi_A$ to $185~\micro\meter$ for $\phi_E$.
Based on Eq. \eqref{eq:resonance}, the observed shift in the resonance  frequency may thus be caused by both the change in $H$ and by a liquid fraction dependency of the sound velocity $c$, whose relative contributions remain to be disentangled.

As a first basic analysis, we measure for all our samples the frequency $\fres{1}$ of the first resonance, as well as the one $\fres{2}$ of the second resonance, when it can be observed.
They are reported in Fig.~\ref{fig:Fres} as a function of the cavity thickness $H$.
For a given resonance peak, all data fall onto a master curve, regardless of the bubble size or liquid fraction.
This suggests that the sound velocity $c$ within the monolayer is independent of these quantities and that the observed shift in the resonance frequency is solely due to the change in $H$.

The solid and dashed red lines in Fig.~\ref{fig:Fres} show fits of the data using the Fabry-P\'erot resonance criterion Eq. \eqref{eq:resonance} for $n=1$ and $2$, respectively. 
The resonance frequencies $\fres{1}$ and $\fres{2}$ convincingly follow a $1/H$ trend and the fitting sound velocities are $207~\meter\per\second$ and $185~\meter\per\second$, respectively.
These values lie very close to the sound velocity in the air and $\perfluoro$ mixture contained within the bubbles, which was independently measured to be $c_g = 184 \pm 7~\meter\per\second$.
This suggests that the non-dissipative propagation of the ultrasonic waves is essentially insensitive to the monolayer liquid fraction and bubble size, but only depends on the gas content of the bubbles.
%
\section{Data analysis and discussion} \label{sec:analysis}
%
\subsection{Porous-like model for data analysis} \label{ssec:sieve}
In our experiments, the acoustic wavelengths in air $\lambda = 200-4000~\micro\meter$ are comparable to the typical length scale of the bubble network $L_s = 150-600~\micro\meter$. 
Therefore, the overall acoustic transmission can be expected to be sensitive to the monolayer structure, even if the resonance frequency itself is not. \bigskip \\
\begin{figure}[t!]
\centering
\includegraphics[width=8cm]{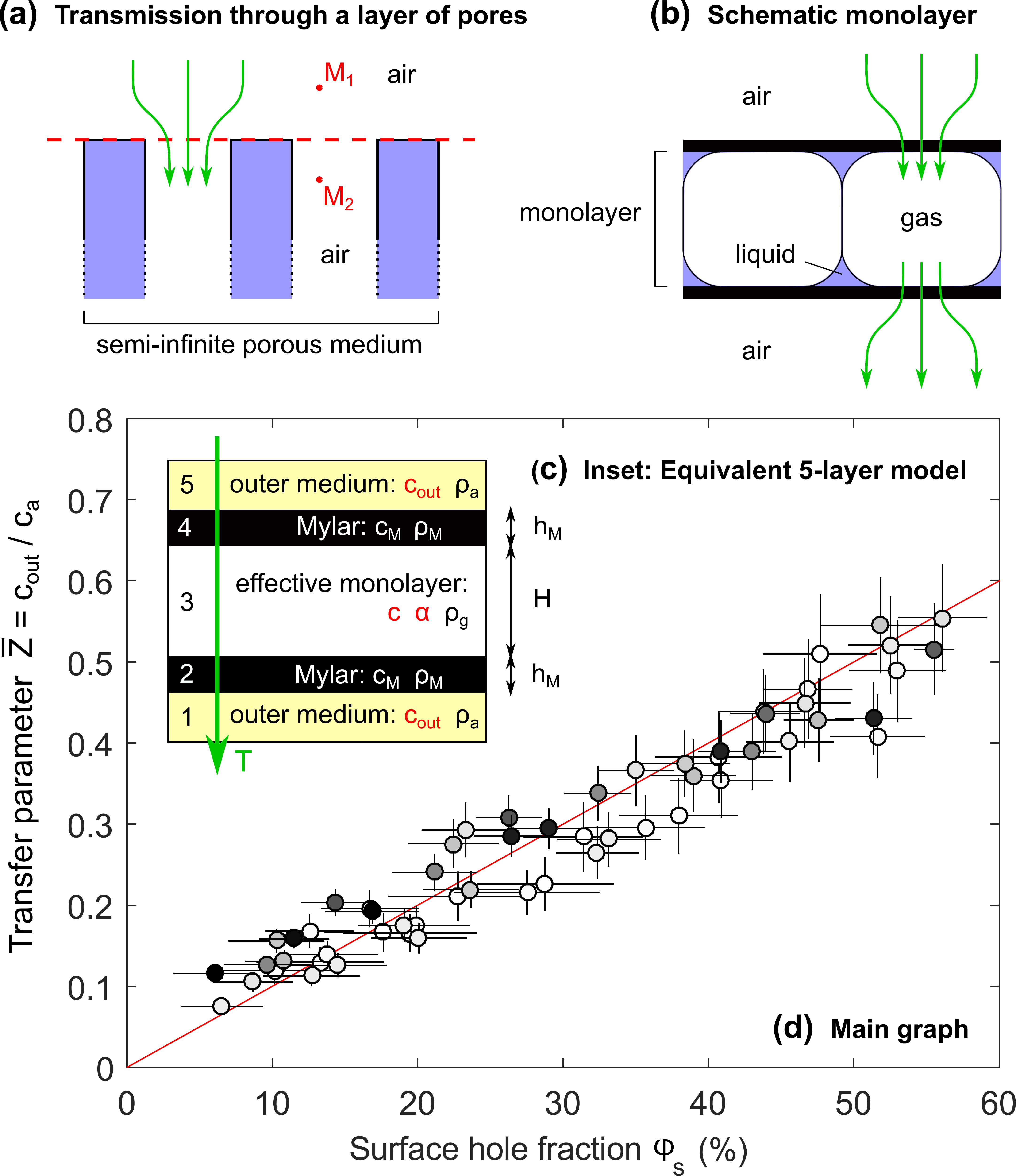}
\caption{(a) Layer of identical parallel pores and acoustic path of a sound wave entering the porous layer (arrows). --- (b) Schematic side view of a monolayer, showing the acoustic path (arrows) avoiding liquid-filled high acoustic impedance zones, as in a porous layer. --- (c) Inset: Stack of 5 layers considered to compute the acoustic transmission $T$. The adjustable parameters of the model are highlighted in red. --- (d) Main graph: transfer parameter $\bar{Z}$ as a function of the monolayer surface hole fraction $\phis$. The solid red line represents the equality $\bar{Z}=\phis$ and the greyscale codes for the bubble volume $V$ (the darker the symbol, the larger the bubble volume).} 
\label{fig:sieve}
\end{figure}
%
\emph{Description of the model.}
In a layer of porous medium with identical parallel pores (see Fig.~\ref{fig:sieve}a), it is known that the acoustic impedance within the pore $Z_2 = Z(M_2)$ is related to the one in free air $Z_1 = Z(M_1)$ through a transfer factor \cite{Allard2009}:
\begin{equation}
Z_2 = \phis \times Z_1,
\label{eq:impedance_porous}
\end{equation}
where $\phis$ is the pore fraction.
This is due to the fact that the velocity field has to adapt when entering a pore in order to satisfy mass conservation, as pictured qualitatively by the green arrows (showing the acoustic path) in Fig.~\ref{fig:sieve}a.
%
%

For an incident ultrasonic wave, the monolayer appears as a regular hexagonal network of low acoustic impedance ``pores'' (the gas phase in bubbles) separated by high acoustic impedance zones (the liquid network).
As sketched in Fig.~\ref{fig:sieve}b, the actual pores in the monolayer have a non-uniform structure in the direction of propagation, due to the curvature of PB and psPB.
As a first approximation, we will not take this detailed shape into account and assimilate the monolayer structure to a hexagonal lattice of identical parallel pores of uniform profile along the direction of propagation.
By analogy with porous materials, we expect this structure to behave as a porous plate through which the acoustic wave has to squeeze in (see Fig.~\ref{fig:sieve}b). 
In reality, our system is more complicated since the pores are enclosed between Mylar sheets, but we will assume that the Mylar membranes locally follow the motion of air and do not affect the porous-like effect. \bigskip \\
%
%
%
\emph{Model implementation.}
In practice, we model our system as a stack of five homogeneous media, represented in Fig.~\ref{fig:sieve}c.
As detailed in the appendix A of reference \cite{Pierre2013}, the transmission $T$ through such a system can be computed knowing (i) the acoustic impedance in layers 1 to 5 and (ii) the wave vector and thickness of layers 2 to 4.
Remarkably, the medium outside the cavity (layers 1 and 5) plays no role in the propagation. It is therefore assigned the same density $\rho_a$ as free air but, as a mathematical trick to incorporate the transfer factor describing the porous-like effect, we consider it has an unknown sound velocity $\cAapp$.
This will allow us to check that the transfer relationship for a porous layer (Eq. \eqref{eq:impedance_porous}) is also valid in our case, namely that $\cAapp = \phis \times \cA$.
Layers 2 and 4 are the Mylar sheets, whose properties (sound velocity $c_M$, density $\rho_M$ and thickness $h_M$) are well-characterized.
Since Fig.~\ref{fig:Fres} tends to show that the acoustic propagation is essentially sensitive to the gas phase contained in the bubbles, the effective monolayer (layer 3) is taken as a homogeneous medium of known density $\rho_g$ and thickness $H$, but adjustable sound velocity $c$ and attenuation $\alpha$.

For each sample, the modulus of the experimental transmission $|T|$ is fitted in the vicinity of the first resonance by the five-layer model described above.
The sound velocity $c$ and attenuation $\alpha$ in the cavity and the sound velocity in the outer medium $\cAapp$ are used as adjustable parameters, yielding the solid red line in the example presented in Fig.~\ref{fig:transmission_mono}a.
The phase of the fitted complex transmission (dashed red line in Fig.~\ref{fig:transmission_mono}a) is also observed to be in fairly good agreement with the experimental measurements.
In the following paragraph, we report and discuss the three fitted parameters as functions of the monolayer characteristics.
%
\subsection{Fitted parameters} \label{ssec:fit_params}
%
\emph{Transfer parameter.}
In our experiments, we define the transfer parameter $\bar{Z}$ as the ratio of the fitted sound velocity in the outer medium $\cAapp$ over the sound velocity in free air $\cA$: 
\begin{equation}
\bar{Z} = \frac{\cAapp}{\cA}.
\label{eq:sieve_param}
\end{equation}
The value of $\cA$ lies in the range $345-348~\meter\per\second$, depending on the environmental conditions on the day of each experiment.
Note that $\bar{Z}$ is equivalently the ratio of the corresponding acoustic impedances.
In analogy with a porous layer, we expect $\bar{Z}$ to be equal to the surface pore fraction $\phis$ \cite{Allard2009}.
The surface pore fraction $\phis$ of a 2D foam, defined as the ratio of the area of the monolayer that is not covered by PB or psPB over the total surface area, is determined independently for each sample using Surface Evolver simulations (see paragraph \ref{ssec:mono_charac}).
In Fig.~\ref{fig:sieve}d (main graph), we plot the transfer parameter $\bar{Z}$ as a function of the surface pore fraction $\varphi_s$.
The experimental data convincingly collapse onto the $\bar{Z} = \phis$ line (plotted in red), thus supporting the consistency of the porous plate analogy.\bigskip \\
\emph{Sound velocity.}
%
The sound velocity $c$ within the effective monolayer enclosed in the cavity is plotted in the inset of Fig.~\ref{fig:Fres} as a function of the volume liquid fraction $\phi$.
This quantity appears to be independent of the liquid fraction, as well as of the bubble volume.
Averaging over all samples, we find a sound velocity $\langle c\rangle = 186 \pm 8~\meter\per\second$.
This value is in excellent agreement with the sound velocity $c_g = 184 \pm 7~\meter\per\second$ measured independently in the air/$\perfluoro$ mixture alone.
This result is consistent with the basic analysis carried out on the resonance frequencies in paragraph \ref{ssec:raw_data} and supports our five layer modeling taking a density within the cavity equal to the one of the gas phase contained in bubbles (see Fig.~\ref{fig:sieve}c). \bigskip \\
%
\emph{Attenuation.}
%
\begin{figure}[t!]
\centering
\includegraphics[width=8cm]{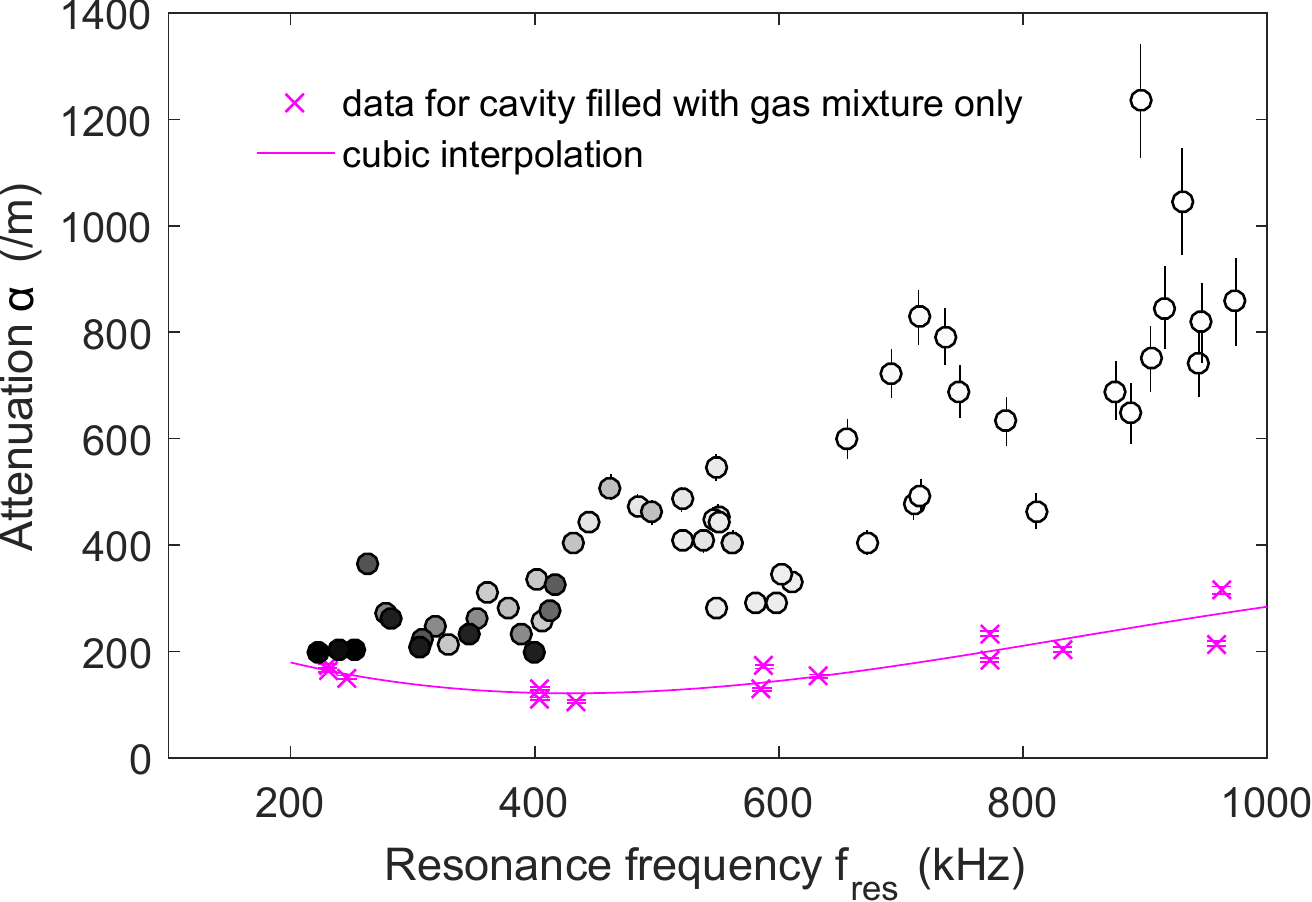}
\caption{Sound attenuation $\alpha$ as a function of the frequency for cavities filled with bubble monolayers ($\circ$) and gas phase only ($\times$). The solid line is a cubic fit of the data for the gas-filled cavity.}
\label{fig:raw_attenuation}
\end{figure}
%
%
The sound attenuation $\alpha$ within the effective monolayer enclosed in the cavity is plotted in Fig.~\ref{fig:raw_attenuation} as a function of the frequency of the first resonance $\fres{1}$ (black circles).
The attenuation is found to increase with frequency but the data points are quite scattered.
This hints at an additional dependency of $\alpha$ with another parameter, which has not been identified yet.
We compare the attenuation in bubble monolayers to the one in a cavity filled only with the gas phase (air and $\perfluoro$ mixture), shown with crosses in Fig.~\ref{fig:raw_attenuation}. 
The attenuation in the gas-filled cavity diminishes when the frequency decreases but, interestingly, it seems to rise again below $400~\kilo\hertz$.
Since this behavior was not observed on non-metalized cavities, for which the attenuation continues to go down towards zero at low frequencies (data not shown), we suppose it is due to the presence of the gold layer coated on half the cavity surface.
For a given frequency, bubble monolayers are observed to attenuate more than the cavity filled with the gas phase only, except at the lowest frequencies.
This suggests that, unlike the sound velocity $c$, the attenuation $\alpha$ within the monolayer is sensitive to the presence of the liquid network as well.
%
\subsection{Discussion on attenuation} \label{ssec:discuss_alpha}
\begin{figure}[t!]
\centering
\includegraphics[width=8cm]{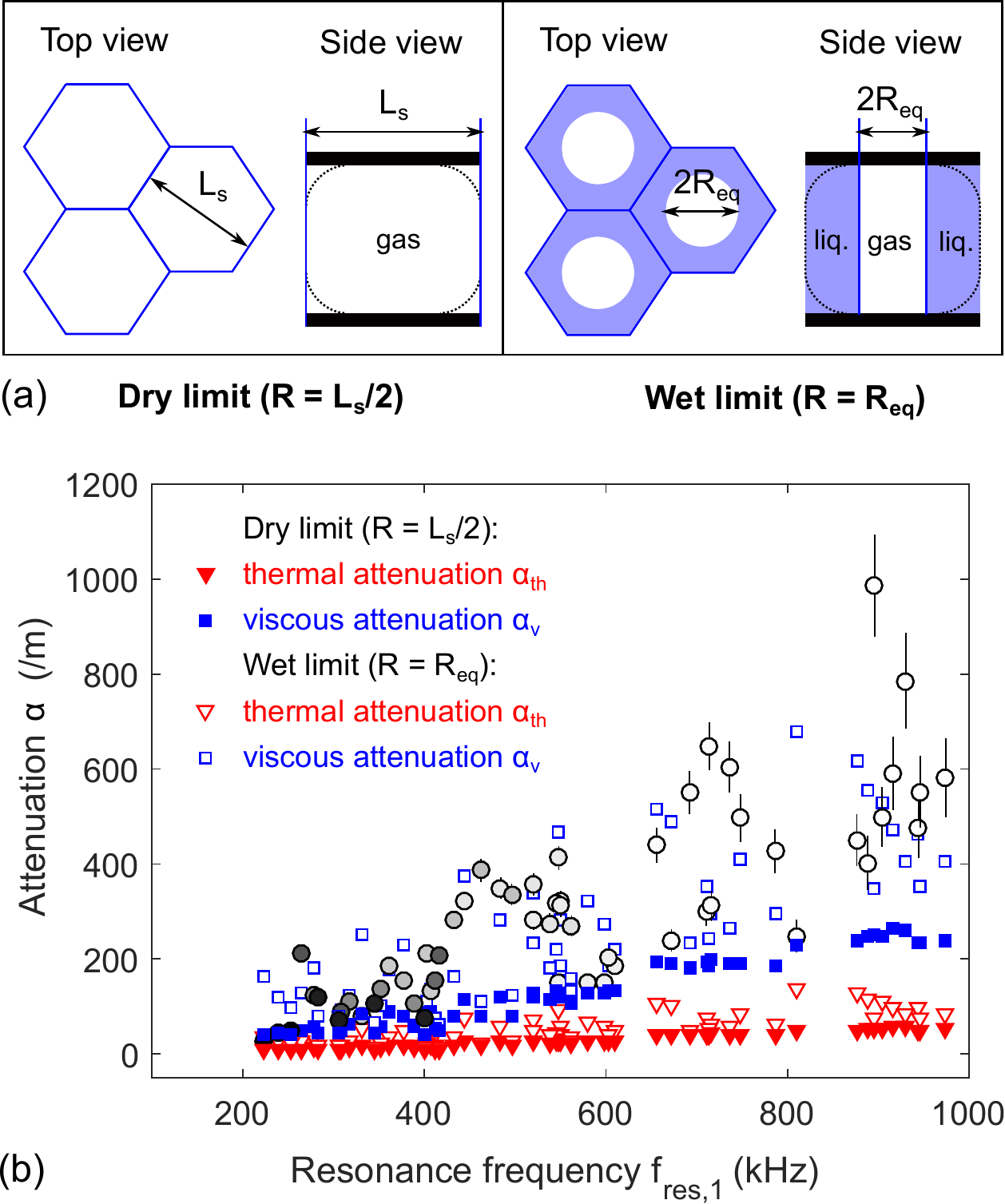}
\caption{(a) Definition of the characteristic pore length scale $R$ in the dry limit ($R = L_s$) and in the wet limit ($R =\Rpore$). --- (b) Attenuation $\alpha$ measured in the monolayer-filled cavity minus the attenuation in the gas-filled cavity, as a function of the resonance frequency $f_{\mathrm{res}}$ (black circles). The greyscale codes for the bubble volume $V$ (the darker the symbol, the larger the bubble volume). The red triangles and blue squares respectively show the contributions expected from thermal and viscous losses, calculated both in the dry limit (filled symbols) and in the wet limit (empty symbols). }
\label{fig:attenuation}
\end{figure}
In order to discuss the possible sources of dissipation in bubble monolayers, we first subtract the attenuation stemming from the cavity itself (solid line in Fig.~\ref{fig:raw_attenuation}) to the attenuation measured in monolayer-filled cavities (circles in Fig.~\ref{fig:raw_attenuation}).
The resulting data points are plotted in Fig.~\ref{fig:attenuation}b as circles again.

In the following, we examine two possible sources of dissipation -- thermal and viscous -- and compute the corresponding contributions to the attenuation.
The first step consists in choosing the relevant bubble ``radius'' (or pore size) $R$ to be compared to the thermal and viscous dissipation length scales.
Several definitions of $R$ can be proposed since the bubbles are non spherical, but we will focus on two limiting situations.
On the one hand, we will look at the case of $R = L_s/2$, which corresponds to an infinitely dry foam since the periodicity $L_s$ is a purely geometrical quantity, independent of the liquid fraction (see Fig~\ref{fig:attenuation}a).
This case will be represented by filled symbols (squares or triangles) in Fig.~\ref{fig:attenuation}b.
On the other hand, we will consider the case where $R = \Rpore$, defined as 
\begin{equation}
\Rpore = \frac{L_s}{2} \times \left(\frac{2 \sqrt{3}}{\pi} \phis \right)^{1/2},
\end{equation}
corresponding to the radius of equivalent circular pores having the same surface area as the actual (hexagonal shaped) pores, as sketched in Fig~\ref{fig:attenuation}a.
In a sense, $\Rpore$ takes into account the liquid contents of the foam in a maximal way.
This case will be represented by empty symbols (squares or triangles) in Fig.~\ref{fig:attenuation}b.\bigskip \\
%
\emph{Thermal dissipation.}
When going through an assembly of bubbles, acoustic waves can loose energy by heating up the gas phase.
Pierre \textit{et al.} \cite{Pierre2017} argued that the calculation of thermal attenuation in bubbly liquids (where bubbles remain independent) \cite{Chapman1971} could be extended to the case of liquid foams (where bubbles are in contact).
As detailed in appendix \ref{apdx:attenuation}, we follow their approach in order to compute thermal losses within our bubble monolayers, both in the dry limit assuming $R = L_s/2$ and in the wet limit assuming $R = \Rpore$. 
The corresponding attenuation $\alphath$ is plotted in Fig.~\ref{fig:attenuation}b with red triangles.
Regardless of the definition used for the bubble radius $R$, thermal losses turn out to contribute only marginally to the total attenuation.
A similar result had already been observed by Pierre \textit{et al.} for 3D polydisperse foam samples \cite{Pierre2017}, while thermal losses had been found dominant in other experiments \cite{Mujica2002,Monloubou2015}. \bigskip \\
%
\emph{Viscous dissipation in gas phase.}
It was shown in paragraph \ref{ssec:fit_params} that the 2D foam structure could be consistently assimilated to a layer of porous medium with a pore fraction equal to the monolayer surface fraction $\phis$ that is not covered by PB or psPB.
In addition to the modification of the gas flow average velocity, the presence of pores may also induce viscous dissipation in a boundary layer close to the pores' lateral walls.
As developed in appendix \ref{apdx:attenuation}, the corresponding attenuation $\alphapore$ can be computed using Kirchhoff's law \cite{Kirchhoff1868,Weston1953,Pierre2017} and is plotted in Fig.~\ref{fig:attenuation}b with blue squares, taking $R = L_s/2$ or $R = \Rpore$.

In the dry limit ($R = L_s/2$, filled squares), the viscous attenuation can account for about half of the total attenuation.
In the wet limit ($R = \Rpore$, empty squares), the pores are narrower due to the presence of thick liquid walls, hence a larger viscous dissipation compared to the dry case: $\alphapore$ becomes comparable to $\alpha$.
This suggests that viscous dissipation in the boundary layer close to the pore walls may be considered a serious candidate to explain ultrasound attenuation in monodisperse bubble monolayers.
However, in order to validate this mechanism, the exact dependencies of $\alphapore$ with frequency $f$ and radius $R$ (see eq. \eqref{eq:alpha_viscous}) should be checked.
Unfortunately, this cannot be done with our experimental configuration in which both the resonance frequency and the bubble radius vary concomitantly, owing to the flexible nature of the cavity.
%
\section{Conclusion} \label{sec:conclusion}
In this work, we investigated the transmission of ultrasound through monodisperse 2D liquid foams generated using microfluidics techniques.
To do so, we developed an original microfluidic cell allowing to enclose a single layer of microbubbles in a cavity of centimetric diameter letting airborne ultrasound through.
A peculiar feature of that setup was the flexibility of the cavity upper and lower walls, thus requiring an independent measurement of the cavity thickness $H$.

The acoustic transmission through monolayers of various bubble sizes and liquid fractions was probed.
The attenuation in the samples was sufficiently small so that an acoustic Fabry-P\'erot resonance could be observed.
A basic analysis of this resonance showed that the sound velocity in monolayers only depends on the gas phase in bubbles and not on the liquid content.
This result was confirmed by a more advanced analysis relying on a five-layers model of system, which was used to fit the acoustic transmission curves in the vicinity of the first resonance.
In this approach, the 2D foam structure perpendicular to the direction of wave propagation was accounted for by a transfer parameter, that was found equal to the monolayer surface pore fraction determined independently using Surface Evolver simulations.
The attenuation within the monolayers was also extracted, and turned out to of the same order as the one expected from viscous losses in the gas phase.
However, this mechanism could not be confirmed due to the simultaneous variation of several parameters, which is intrinsic to our experimental setup.

Future experiments will look into the acoustic transmission through several layers of monodisperse bubbles, for which a qualitatively different behavior is expected, due to the presence of free liquid films \cite{Pierre2014}.
%
\section*{Acknowledgments}
L.C. and M.C.J. acknowledge funding from Agence Nationale de la Recherche (contract number 13-BS09-0011-01) and PSL Research University. This work has received the support of Institut Pierre-Gilles de Gennes (laboratoire d'excellence, ``Investissements d'avenir'' program ANR-10-IDEX-0001-02 PSL and ANR-10-LABX-31; \'equipement d'excellence, ``Investissements d'avenir'', program ANR-10-EQPX-34). 

We are very grateful to Guillaume Lafitte, Olivier Lesage and Nawel Cherkaoui from the IPGG platform for their technical support. Many thanks to \'Elian Martin, Jules Dupire and Alexandre Mansur for valuable help regarding the experiments. We are also indebted to Cyprien Gay and Albane Thery for fruitful discussions regarding Surface Evolver simulations.
%
\section*{Author contribution statement}
LC, JP, VL and MCJ designed the experimental setup. LC performed the experimental measurements, which were analyzed by LC an JP. The Surface Evolver simulations were done by AD. All authors discussed the results, interpreted them and were involved in the preparation of the manuscript, which was mainly written by LC.
%
\appendix
\section{Topography of the cavity} \label{apdx:profile}
\begin{figure}[t!]
\centering
\includegraphics[width=8cm]{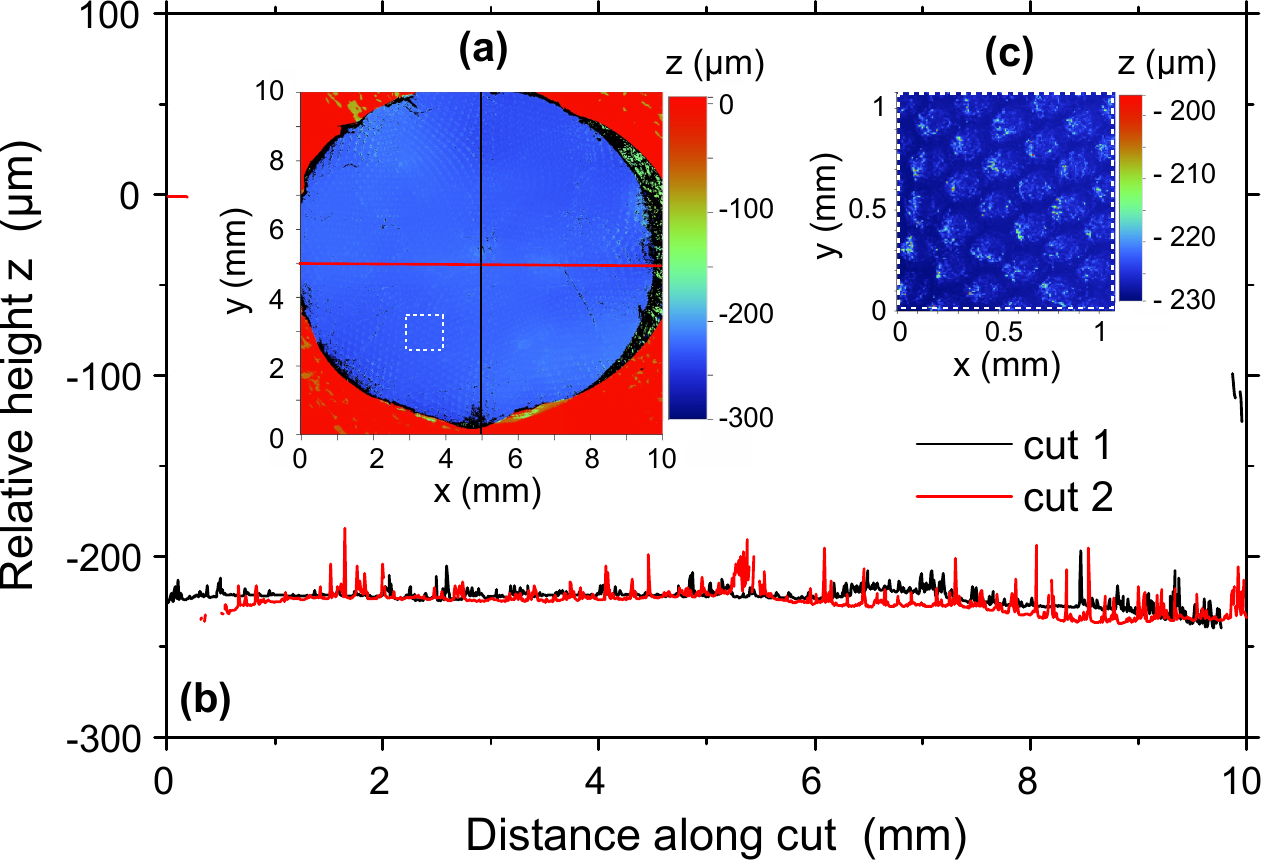}
\caption{(a) Two-dimensional map of the relative height $z$ between the top glass slide ($z=0$) and Mylar membrane for a cavity filled with a monodisperse bubble monolayer. --- (b) One-dimensional height profiles along the two cuts marked on the map. --- (c) Close-up on the zone situated in the white square on the map. Note that the color code has been re-adjusted for better contrast.}
\label{fig:profilo_opt}
\end{figure}
The cavity in which the monolayers are enclosed is sealed by two $1~\micro\meter$-thick Mylar membranes that may deform upon bubble injection.
Since the thickness $H$ of the cavity is a crucial parameter in our data analysis, we investigate in this appendix the potential heterogeneity of $H$ over the whole cavity surface.
To do so, we inject a monodisperse bubble monolayer in one of our microfluidic cells with a cavity diameter $D = 10~\milli\meter$.
Note that the diameter of this cavity is a little smaller than in the ones used for acoustic measurements ($D = 16~\milli\meter$), so that the whole surface topography may be scanned before the foam starts coarsening.

Using an optical profilometer (NT9100, Veeco), the height $z$ of the top Mylar membrane relatively to the top glass slide ($z=0$) is measured in each point of the cavity surface.
The resulting 2D topography map is presented in Fig.~\ref{fig:profilo_opt}a.
Cuts along two orthogonal diameters of the cavity are extracted and displayed in the main graph (Fig.~\ref{fig:profilo_opt}b).
These data show that the cavity filled with a bubble monolayer exhibits no large-scale deformation.
Interestingly, the close-up presented in Fig.~\ref{fig:profilo_opt}c (corresponding to the white square area marked in Fig.~\ref{fig:profilo_opt}a) reveals local deformations due to the bubble lattice enclosed.
However, the amplitude of these local bubble-induced perturbations remains smaller than $10~\micro\meter$, which is the typical uncertainty on the cavity thickness measurement (see paragraph \ref{ssec:mono_charac}).
The cavity thickness $H$ can thus be safely regarded as a homogeneous quantity, as done throughout this study.
\section{Expressions for thermal and viscous attenuations} \label{apdx:attenuation}
%
In this appendix, we give the expressions for thermal and viscous attenuation that we used in order to obtain the different contributions presented in Fig.~\ref{fig:attenuation}b. Note that these expressions will be evaluated using two different definitions of the bubble ``radius'' $R$ in our 2D foams : $R = L_s/2$ for the dry limit and $R = \Rpore$ for the wet limit. \bigskip \\
%
\emph{Thermal dissipation.}
Following the arguments by Pierre et al. \cite{Pierre2017}, we consider that thermal attenuation in a 2D foam has the same expression as in a bubbly liquid \cite{Chapman1971}, namely 
\begin{equation}
\alphath = - \frac{\mathrm{Im} (\kappa)}{2 \, \mathrm{Re} (\kappa)} \times \frac{\omega}{c}
\label{eq:alpha_thermal}
\end{equation}
where $\omega$ is the angular frequency and $\kappa$ the complex polytropic exponent defined as
\begin{equation}
\kappa = \frac{\gamma}{1 + 3(\gamma - 1) (1 - X \, \mathrm{cotan} \, X)/ X^2 }.
\end{equation}
In this expression, $X = (1 + \mathrm{i}) \, R / \ell_{\mathrm{th}}$ is the ratio between the bubble radius $R$ and the thermal length $\lth = \sqrt{2 D_{\mathrm{th}}/\omega}$, $D_{\mathrm{th}}$ and $\gamma$ are respectively the thermal diffusivity and ratio of heat capacities in the gas phase.
The two latter parameters were estimated for the air/$\perfluoro$ mixture used in the experiments as explained in reference \cite{Pierre2017}, and were found to be $D_{\mathrm{th}} = 7.5~\milli\meter^2\per\second$ and $\gamma = 1.12$. \bigskip \\
%
%
\emph{Viscous dissipation in gas phase.}
Assimilating the bubbles in the monolayer to parallel cylindrical pores of radius $R$, dissipation can take place in the gas phase over a typical viscous length $\lv = \sqrt{2 \eta_g / \rho_g \omega}$, where $\rho_g$ and $\eta_g$ are the gas density and viscosity, respectively.
The corresponding viscous attenuation $\alphapore$ is then given by Kirchhoff's law \cite{Kirchhoff1868,Weston1953}
\begin{equation}
\alphapore = \frac{\omega}{c} \times \frac{\ell_v}{2R}%
= \frac{1}{2 c} \, \sqrt{\frac{2 \eta_g \omega}{\rho_g R^2}}.
\label{eq:alpha_viscous}
\end{equation}
\bibliography{biblio}
\bibliographystyle{unsrt}
%
\end{document}